\documentclass[twocolumn,preprintnumbers,amsmath,amssymb]{revtex4}
\usepackage{amsmath,amssymb,graphics,epsfig,subfigure}
\usepackage{color}

\begin{document}

\thispagestyle{empty}

\begin{center}

\title{Logarithmic corrections to black hole entropy and holography}

\date{\today}
\author{Aritra Ghosh\footnote{E-mail: ag34@iitbbs.ac.in}}

\affiliation{School of Basic Sciences, Indian Institute of Technology Bhubaneswar,\\   Jatni, Khurda, Odisha, 752050, India}

\author{Chandrasekhar Bhamidipati\footnote{E-mail: chandrasekhar@iitbbs.ac.in}}

\affiliation{School of Basic Sciences, Indian Institute of Technology Bhubaneswar,\\   Jatni, Khurda, Odisha, 752050, India}

\author{Sudipta Mukherji\footnote{E-mail: mukherji@iopb.res.in}}

\affiliation{Institute of Physics, Sachivalaya Marg, Bhubaneswar, 751005, India\\}

\affiliation{Homi Bhabha National Institute, Training School Complex, Anushakti Nagar, Mumbai, 400085, India}

\begin{abstract}
We compute logarithmic corrections to the black hole entropy $S_{\rm bh}$ in a holographic set up where the cosmological constant $\Lambda$ and Newton's constant $G_D$ are taken to be thermodynamic parameters, related to variations in bulk pressure \(P\) and central charge \(c\). In the bulk, the logarithmic corrections are of the form: $\mathcal{S} = S_{\rm bh} - k \ln S_{\rm bh} + \cdots$ arising due to fluctuations in thermodynamic volume, induced by a variable $\Lambda$, in addition to energy fluctuations. We explicitly compute this coefficient $k$ for the BTZ black hole and show that the result matches with the one coming from the logarithmic corrections to the Cardy's formula. We propose an entropy function in the CFT, which exactly reproduces the logarithmic corrections to black hole entropy in arbitrary dimensions.
\end{abstract}
\maketitle
\end{center}

\textit{Introduction -}
Bekenstein-Hawking entropy of black holes~\cite{Bekenstein:1973ur,Hawking:1974rv} 
\begin{equation}\label{sbh}
S_{\rm bh} = \frac{A}{4G_D}\, ,
\end{equation} 
and its statistical interpretation~\cite{Vafa,Maldacena,Horowitz,Emparan,Mathur,Carlip,1109.3706,0002040,0411035,0905.3168,parthasarathi,Mukherji:2002de,Sen:2012dw} is important to learn about the degrees of freedom of quantum gravity. Here,  $A$ is area of the horizon, $G_D$ is the gravitational Newton's constant in \(D\) dimensions, and we set $k_B=\hbar=c=1$. In $(2+1)$-dimensions, Cardy's formula for asymptotical growth of states~\cite{Cardy:1986ie},
\begin{equation}
\label{cardy}
S_{\rm Cardy}
= 4 \pi \sqrt{ {c\over 24}\left(\Delta - {c\over 24}\right) } \, 
+ 4 \pi \sqrt{ {\bar c\over 24}\left(\bar \Delta - {\bar c\over 24}\right) } 
\end{equation}
matches $S_{\rm bh}$ for BTZ black holes~\cite{Brown:1986nw,Banados:1992wn,Strominger:1997eq}, where $c = \bar c= 3l/2G_3$, 
with $l$ and $G_3$ being the AdS radius and Newton's constant in three dimensions respectively. The conformal dimensions \((\Delta, \bar{\Delta})\) are related to the black hole parameters as
\begin{equation}\label{delta}
\Delta = \frac{(r_++r_-)^2}{16 G_3 l}, \, \qquad \bar\Delta= \frac{(r_+-r_-)^2}{16 G_3 l}\, .
\end{equation}
Here $r_+,r_-$ are the outer and inner horizons of the BTZ black hole. States with large dimension in the conformal field theory (CFT) are regarded as microstates of the BTZ black hole. The logarithmic corrections to the Cardy's formula also reproduce the first order quantum corrections to black hole entropy, whose generic form is:
\begin{equation} \label{log}
\mathcal{S} = S_{\rm bh} - k \ln S_{\rm bh} + \cdots\, .
\end{equation} 
The coefficient \(k\) varies depending on the number of dimensions, the specific gravity theory etc.~\cite{Sen:2012dw}.\\

\noindent
It is now possible to study the effect of volume fluctuations on logarithmic corrections to black hole entropy in eqn (\ref{log})~\cite{Ghosh:2021uxg,Ghosh:2021upm}, using the extended thermodynamic set up in AdS backgrounds~\cite{Kastor:2009wy,Dolan:2010ha,Cvetic:2010jb,Kubiznak:2012wp,K,Johnson:2018amj,Kastor:2010gq,visser,mann,Johnson,K1,K2}, where a variable cosmological constant $\Lambda$ gives thermodynamic pressure \cite{Kastor:2009wy}:
\begin{equation}\label{P}
P=-\frac{\Lambda}{8\pi G_D}\,,\quad \Lambda=-\frac{(D-1)(D-2)}{2 l^2}\,.
\end{equation}
Infact, holographic applications of extended thermodynamic are now conceivable due to 
a new mixed form of the first law of black hole mechanics proposed in~\cite{mann}:
\begin{equation}
\delta \mathcal{M} = T_h \delta S_{\rm bh}+V_c \delta P+\mu \delta c\,, \label{FirstC} 
\end{equation}
where $\mathcal{M}$ is the ADM mass, $T_h$ is the Hawking temperature, $\mu$ is the chemical potential conjugate to  the central charge 
\begin{equation} \label{C}
c = j \frac{l^{D-2}}{16 \pi G_D}\, ,
\end{equation}
($j$ depends on the particulars of the theory) and $P$ is the pressure defined in eqn (\ref{P}). Its conjugate is the thermodynamic volume \cite{mann}
\begin{equation}\label{VC}
V_c :=\Bigl(\frac{\partial \mathcal{M}}{\partial P}\Bigr)_{S_{\rm bh},c}\, =\frac{\mathcal{M}}{DP}\, .
\end{equation} The key point is that in eqns (\ref{sbh}), (\ref{P}) and (\ref{C}), the Newton's constant \(G_D\) as well as \(l\) are thermodynamic parameters \cite{visser,mann} (see also \cite{Johnson,K1,K2}). The variations of $G_D$ do not appear explicitly in eqn (\ref{FirstC}), as they have been written in terms of variations of $c$ and $P$ from their definitions.\\ 


\noindent 
Our aim is to use the thermodynamic quantities appearing in the first law [eqn (\ref{FirstC})] and compute the leading logarithmic corrections to the black hole entropy, following the methods developed in~\cite{Ghosh:2021uxg,Ghosh:2021upm}. We also compare our results with corrections to Cardy's formula from the corresponding boundary CFT. In the bulk, corrections to entropy in eqn (\ref{log}) come from energy fluctuations and additionally those of the volume $V_c$ induced by the variable cosmological constant $\Lambda$. On the boundary, logarithmic corrections to the Cardy's formula can be computed directly by exploiting the modular invariance of the partition function \cite{Strominger:1997eq,Carlip}. A generic entropy function on the boundary reproduces the logarithmic entropy corrections consistent with bulk expectation. Although, our key results are expected to be valid in general dimensions, the connection of Cardy's formula to black hole entropy is much clearer in three dimensional gravity, and we concentrate on this case.\\

\noindent
\textit{Bulk -} 
In the extended thermodynamics set up, since \(\mathcal{M}\) can be interpreted as the enthalpy (rather than internal energy), the relevant free energy is the Gibbs potential \(\mathcal{G} = \mathcal{M} - T_hS_{\rm bh}\) whose variations satisfy
\begin{equation}
\delta \mathcal{G} = -S_{\rm bh} \delta T_h + V_c\delta P + \mu \delta c .
\end{equation}
The equilibrium state of the system (here the black hole) is dictated by the Gibbs potential, and thus, it is natural to choose the isothermal-isobaric ensemble, where intensive variables \(T\) and \(P\) are fixed by respective reservoirs \footnote{The central charge \(c\) is fixed at a particular equilibrium state although it may in principle vary from one equilibrium to another.}. Consequently at finite temperature, there are thermodynamic fluctuations in energy and volume around equilibrium, which alter the microscopic density of states and lead to corrections to black hole entropy.\\

\noindent
For any thermodynamic system in the isothermal-isobaric ensemble, the partition function is
\begin{equation}\label{lt}
  \Delta(\beta,\beta P) = \Gamma \int_{0}^{\infty} \int_{0}^{\infty} \Omega(E,V) e^{-\beta (E + PV)} dE dV.
\end{equation}  The pre-factor \(\Gamma\) has dimensions of inverse volume which is necessary to make \(\Delta(\beta,\beta P)\) dimensionless and \(\Omega(E,V)\) is the density of states. These Laplace transforms can be inverted to give 
 \begin{eqnarray}
    \Omega(E,V) &=& \frac{\Gamma^{-1}}{(2 \pi i)^2} \oint \oint \Delta(\beta,\beta P) e^{\beta (E + PV)} d\beta d(\beta P) \nonumber \\
     &=& \frac{\Gamma^{-1}}{(2 \pi i)^2} \oint \oint e^{S(\beta, \beta P)} d\beta d(\beta P).
  \end{eqnarray}
To obtain the second equality, we have used \(\Delta(\beta,\beta P) = \exp (-\beta \mathcal{G})\) where \(\mathcal{G} = E + PV - TS\) is the Gibbs free energy. The integrals above, can be evaluated by saddle point method by expanding \(S(\beta, \beta P)\) about equilibrium up to the second order. If \(\Omega_0 = e^{S_0}\) be the number of microstates at equilibrium, then the microcanonical entropy \(\mathcal{S} = \ln \Omega\) gets log corrected and can be computed to be~\cite{Ghosh:2021uxg,Ghosh:2021upm}
\begin{equation}\label{logcorrectiongeneral}
\mathcal{S} = S_0 - \frac{1}{2} \ln (\Gamma^{-1} T^3 V \kappa_T C_V ) + \cdots \, .
\end{equation} 
The above result is rather general and holds for any system whose statistical mechanics is described by the isothermal-isobaric ensemble. For a black hole, we make the identifications \(S_0 \rightarrow S_{\rm bh}\), \(V \rightarrow V_c\), \(T \rightarrow T_h\) and we choose $\Gamma=G_3^2$ to fix the scale. \\

\noindent
Now, consider the BTZ black hole. For brevity, we consider the zero angular momentum (\(r_- = 0\)) case, although our results are valid for non-zero angular momentum at high temperatures (\(r_+\) large). The mass (enthalpy), \(S_{\rm bh}\) and the Hawking temperature \(T_h\) are
\begin{equation} \label{SPC}
\mathcal{M} = \frac{r_+^2}{2 G_3 l^2} \, , \hspace{4mm} S_{\rm bh} = \frac{\pi r_+}{2 G_3}, \hspace{4mm} 
T_h = \frac{G_3 S_{\rm bh}}{\pi ^2 l^2} \, ,
\end{equation} 
and from eqn (\ref{VC}), volume is $V_c = \frac{4 G_3^2 S_{\rm bh}^2}{3 \pi }$.
Second order response functions which we require read
\begin{equation}
C_{V_c} = \frac{S_{\rm bh}}{2} \,, \hspace{5mm} \kappa_T= \frac{3}{4P} \,.
\end{equation}
Subsequently, eqn (\ref{logcorrectiongeneral}) gives the leading logarithmic corrections to the black hole entropy as
\begin{equation}\label{BTZbulk}
\mathcal{S} = S_{\rm bh} - 3 \ln S_{\rm bh} + 2\, \ln c \, + \cdots\, ,
\end{equation} 
where eqns (\ref{P}) and (\ref{C}) have been used as required, and some constants are disregarded. These are the logarithmic corrections to the entropy of the BTZ black hole due to fluctuations in energy, i.e. \(E = \mathcal{M} - P V_c\) and volume \(V_c\).\\

\noindent
\textit{Boundary-} Logarithmic corrections to the Cardy's formula in eqn (\ref{cardy}) can be computed in a two dimensional conformal field theory and are known to be of the following form~\cite{Carlip}:  
\begin{equation} \label{bdy}
\mathcal{S} = S_{\rm bh} - 3 \ln S_{\rm bh} + 2\, \ln c + \cdots\, ,
\end{equation} 
where in the above equation, we included additional factors of $2$ compared to~\cite{Carlip}, as we
take the contribution of both left and right moving CFTs, and \(r_- = 0\) making \(\Delta = \bar{\Delta}\) for the non-rotating black hole. This is in perfect agreement with the result from eqn (\ref{BTZbulk}) in the bulk. In particular, the coefficient of the leading logarithmic terms in eqn (\ref{bdy}) matches that obtained in the bulk and can in fact also be obtained by postulating a boundary entropy function. The left and right movers in the CFT are associated with inverse temperature parameters \(\beta_L\) and \(\beta_R\) respectively, from modular symmetry, it is suggestive to consider the following ansatz~\cite{parthasarathi,Hartman:2014oaa} for an exact entropy function:
\begin{equation}\label{entropyfunction}
S(\beta_L,\beta_R) = a ( \beta_L + \beta_R) + b \bigg( \frac{1}{\beta_L} + \frac{1}{\beta_R} \bigg)
\end{equation} where \(a\) and \(b\) are suitable positive constants. Although, \(\beta_L\) and \(\beta_R\) are independent parameters, for the neutral and non-rotating black hole, at equilibrium, one has \(\beta_0 = (\beta_L)_0 = (\beta_R)_0\) and \(S((\beta_L)_0 , (\beta_R)_0) := S_0\), where \(\beta_0\) is proportional to inverse Hawking temperature.  In the saddle point approximation, the logarithmic corrections to microcanonical entropy are given by \cite{Ghosh:2021uxg}
\begin{equation} \label{Sbulk}
\mathcal{S} = S_0 - \frac{1}{2} \ln D_0 + \cdots
\end{equation} where \(D_0\) is the determinant of the matrix of second derivatives (the Hessian) evaluated at \(((\beta_L)_0,(\beta_R)_{0})\). A straightforward computation reveals that \(D_0 = S_0^2/4 (\beta_L)_0^2 (\beta_R)_0^2\) giving
\begin{equation} \label{SCFT}
\mathcal{S} = S_0 -  \ln S_0 T_0^2 + \cdots
\end{equation} where we have used the fact that \(1/T_0 = \beta_0 = (\beta_L)_0 = (\beta_R)_0\). For the BTZ black hole, identifying \(S_0 \sim S_{\rm bh}\) and \(T_h \sim \, T_0/l\) together with the expression for Hawking temperature given in eqns (\ref{SPC}), gives eqn (\ref{bdy}) exactly, reproducing not only the coefficient \(k = 3\), but also the term \(2 \ln c\) found in the bulk. \\

\noindent
Now, the boundary computations given in eqns (\ref{bdy}) and (\ref{SCFT}) are inherently valid in the Cardy regime:
\begin{equation} \label{carregime}
\Delta/ c \to \infty, {\ c\ } ~({\rm fixed}) \, ,
\end{equation} 
whereas the result in eqn (\ref{logcorrectiongeneral}) is valid in the semi-classical black hole regime:
\begin{equation} \label{bhregime}
c\to \infty, {\ \Delta/c\ }~({\rm fixed}) \, .
\end{equation}
There have been interesting attempts to reconcile the two regimes~\cite{Witten:2007kt,Maloney:2007ud,Heemskerk:2009pn,El-Showk:2011yvt,Keller:2011xi,Hartman:2014oaa,Belin:2014fna,DeLange:2018wbz}. One of the ways is to consider
an $N$-fold symmetric product of certain ``seed" CFTs, such that each CFT has a fixed central charge. The total central charge of the boundary is then $c=24\,N$, though $N$ can vary. The partition function of the  product CFT can be computed in the saddle point approximation and the region of validity of the Cardy's formula can be shown to be naturally extendable to the black hole regime~\cite{DeLange:2018wbz}. We however have to encounter a scenario where $c$ is, in principle a variable (though the central charge of individual CFTs is fixed). This is precisely the holographic black hole chemistry set up [eqn (\ref{FirstC})] which takes into account variations in central charge in the thermodynamic description of bulk geometry. \\

\noindent
We thus check the validity of our result in eqn (\ref{logcorrectiongeneral}), by computing the logarithmic corrections to the Cardy's formula when the boundary has a product CFT structure~\cite{DeLange:2018wbz}. Here, one starts by considering a family of chiral CFTs (labeled by $N$) with large central charge $c=24\, N$ such that $c\to\infty$ as $N\to\infty$.  
For each seed CFT, the partition function is~\cite{DeLange:2018wbz}
\begin{equation}
Z_N(\tau) \equiv {\rm Tr}~ q^{(\Delta-c/24)}
\qquad \qquad q=e^{2\pi i\tau} \, ,
\end{equation}
where the trace runs over the Hilbert space of the $N^{th}$ conformal field theory. Here, ${\rm Im}~\tau$ plays the role of  the inverse temperature and  $Z_N(\tau)$ is the usual canonical partition function. With $\tau$ as the modular parameter of the torus, $Z_N(\tau)$ is invariant under the standard $SL(2,Z)$ transformations which leads to the Cardy's formula in eqn (\ref{cardy}), in the limit when $N$ is fixed. When $N$ is not fixed, one can proceed by introducing a second modular parameter $\rho$, whose imaginary part represents the chemical potential $\mu$, dual to the integer $N$. The partition function motivated from the DMVV~\cite{Dijkgraaf:1996xw} product formula is~\cite{DeLange:2018wbz}
\begin{equation}\label{grand1}
{\cal Z}(p,q) =\sum_{N=0}^\infty p^{N+1} Z_N(q) =  \sum_{N>0, M>-N} q^M p^N \Omega(M,N) \, ,
\end{equation}
where \(q = e^{2 \pi i \tau}\) and \(p = e^{2 \pi i \rho}\), while, \(M = \Delta - \frac{c}{24}\) is the shifted conformal dimension. \(\Omega(M,N)\) is the degeneracy which can be expressed by inverting eqn (\ref{grand1}) to give
\begin{eqnarray}
\Omega &=& \frac{1}{(2 \pi i)^2} \oint \oint \mathcal{Z}(q,p) \frac{dq}{q^{M+1}} \frac{dq}{p^{N+1}} \nonumber \\
            &=& \oint \oint \mathcal{Z}(\tau,\rho) e^{-2 \pi i \tau} e^{-2 \pi i \rho} d\tau d\rho.
\end{eqnarray} 
${\cal Z}(p,q)$ permits a ``dual" modular symmetry, under which $\rho$ transforms under $SL(2,Z)$ too. The $\tau \to -1/\tau$ ($\tau \sim \beta$) symmetry relates high and low temperatures, and in an identical fashion, the transformation $\rho \to -1/\rho$ ($\rho \sim \mu$) relates small and large values of $N$~\cite{DeLange:2018wbz}. Disregarding the ground state contributions, the integral can be evaluated by the method of steepest descent in the limit of large $N,M$, giving (see for example \cite{Carlip,DeLange:2018wbz}) 
\begin{eqnarray}
\Omega(N,M) \approx \left({N\over 4M^3}\right)^{1/4}\,
   \exp\left\{ 4\pi\sqrt{NM}\right\} \, ,
\label{b9}
\end{eqnarray}
Using eqns (\ref{delta}) and (\ref{C}), we get
\begin{equation} \label{density}
\Omega \approx \frac{2\sqrt{2\,G_3} l}{r_+^{3/2}} e^{S_{\rm cardy}}.
\end{equation}
Note that the density of states only includes contributions from one sector (since the CFT is chiral) and the argument in the exponential is simply first term in eqn (\ref{cardy}). Assuming possible extension of the arguments presented in~\cite{DeLange:2018wbz} to include contributions from the non-chiral sector too (supposing holomorphic factorisation~\cite{Witten:2007kt}), 
one gets the same corrections, i.e. $k=3$ as obtained from eqn (\ref{bdy}) in exact agreement with the bulk result given in eqn (\ref{logcorrectiongeneral}). \\

\noindent
\textit{\(Generalizations~to~D > 3\) -} The above arguments can be generalized to an arbitrary number of dimensions. For brevity, let us consider the high temperature limit, where it is straightforward to show that the mass, temperature and volume scale with entropy as \(\mathcal{M} \sim S_{\rm bh}^{(D-1)/(D-2)}\), \(T_h \sim S_{\rm bh}^{1/(D-2)}\) and \(V_c \sim S_{\rm bh}^{(D-1)/(D-2)}\) respectively. Using these quantities in eqn (\ref{logcorrectiongeneral}) gives
\begin{equation}\label{logcorrectiongeneral2}
\mathcal{S} = S_{\rm bh} - \frac{D}{D-2} \ln S_{\rm bh}  + \cdots \, .
\end{equation} 
On the boundary, upon identifying \(T_0\) and \(S_0\) appearing in eqn (\ref{SCFT}) with the Hawking temperature \(T_h\) (up to a factor) and Bekenstein-Hawking entropy \(S_{\rm bh}\), and using the scaling \(T_h \sim S_{\rm bh}^{1/(D-2)}\), eqn (\ref{SCFT})  gives the leading logarithmic corrections identical to the bulk result in eqn (\ref{logcorrectiongeneral2}). \\

\noindent
\textit{Discussion -} Let us understand the picture that now emerges in the holographic black hole chemistry set up~\cite{Kastor:2009wy,Johnson:2018amj,visser,mann,Johnson,K1,K2,K}. From eqn (\ref{C}), a variable cosmological constant $l$ (together with a variable $G_D$) corresponds to a varying central $c$ of the CFT.  Although the first law in eqn (\ref{FirstC}) was derived in a variable $c$ set up, the motivation was to eventually study fixed $c$ CFT, while at the same time leading to a variable cosmological constant scenario in the bulk (to have the necessary $P \delta V_c$ or $V_c \delta P$ terms)~\cite{mann}. In this regard, we performed a computation of the logarithmic corrections to the black hole entropy in the bulk, due to the presence of the pressure-volume terms in eqn (\ref{FirstC}) which leads to the result in eqn (\ref{logcorrectiongeneral}). On the boundary, we found perfect agreement in the coefficient of the leading logarithmic corrections to the Cardy formula, and also from an independent ansatz for the exact entropy function motivated by the modular invariance of the partition function. When the boundary is considered to have a product CFT structure, it may be possible to generalise the entropy function in eqn (\ref{entropyfunction}) to include contributions from the chemical potential. One possible ansatz is
\begin{equation}\label{exact2}
S(\beta_L,\beta_R,\mu_L,\mu_R\,) =  \sum_{i=L,R}a \left(\beta_i  \mu_i +\frac{\beta_i }{\mu_i } \right)+ b \left(\frac{\mu_i }{\beta_i }+\frac{1}{\beta_i  \mu_i } \right)
\end{equation} for some (positive) constants \(a\), \(b\). Following the methods discussed earlier, the corrections are of the form in eqn (\ref{Sbulk}), with $D = \frac{S_0^2}{4 \beta_0} \sim S_0^2 T_0$, giving the leading coefficient of the logarithmic corrections to entropy to be $k=3$ once again.

\noindent
If $G_D$ is held fixed, then eqn (\ref{FirstC}) reduces to $\delta M=T_h \delta S_{\rm bh} +V\delta P$~\cite{mann}, where $V$~\cite{Dolan:2010ha, Cvetic:2010jb} differs from $V_c$ significantly, and the logarithmic corrections computed in this scenario~\cite{Ghosh:2021uxg,Ghosh:2021upm} do not match the boundary results any more. The computations performed here need to be understood in better light, as the status of product CFTs as dual to pure AdS gravity is work in progress~~\cite{Witten:2007kt}-\cite{Gaberdiel:2020ycd}.  However, as our results are derived in the limit of large central charge, a single or a product CFT structure fits harmoniously and gives the correct asymptotic growth of states required for the match.  

\noindent
\textbf{Acknowledgements -} A.G. gratefully acknowledges the financial support received from the M.H.R.D., Government of India in the form of a Prime Minister's Research Fellowship. C.B. thanks the SERB (DST), GoI, through MATRICS  grant no. MTR/2020/000135.

\end{document}